
\magnification=1500
\bigskip
\centerline{\bf  HAS A STANDARD PHYSICS SOLUTION TO THE SOLAR}
\smallskip
\centerline{\bf  NEUTRINO PROBLEM BEEN FOUND? - A RESPONSE }

\bigskip
\centerline{\bf Arnon Dar and Giora Shaviv}
\smallskip
\centerline{Department of Physics }
\centerline{and}
\centerline{Asher Space Research Institute}
\centerline{Israel Institute of Technology}
\centerline{Haifa, Israel 32,200}
\bigskip
\noindent
 {\bf Abstract}
\smallskip
\noindent
In a recent paper Dar and Shaviv (1994) presented results of
calculations with an improved standard solar model (SSM) that
suggest a standard physics solution to the solar neutrino problem.
In a subsequent publication with a similar title  Bahcall et al.
(1994) have claimed to refute the results obtained by Dar and
Shaviv.
Here we show that the criticism and conclusions of Bahcall et al.
are unjustified. In particular, their attempts to reproduce the
results of Dar and Shaviv failed because they did not use the same
nuclear cross sections and did not include in
their calculations major physics improvements which were included
in the improved SSM code used by Dar and Shaviv.
\bigskip
\noindent
{\bf Introduction}
\smallskip
\noindent
To facilitate understanding our response to a critique on our paper
"A Standard Physics Solution To The Solar Neutrino Problem ?"
(Dar and Shaviv, Phys. Rev. Lett. Submitted 1994), posted by
15  authors! (Bahcall et al. 1994) on this  electronic bulletin
board, we quote here (in italics) the main comments and criticism
of these authors and
insert our responses (in Roman fonts) between them.
\bigskip
\noindent
 $ \underline{ Bahcall \ et \ al.~abstract:}$
\smallskip
\noindent
{\it "The claim by Dar and Shaviv that they have found a standard
model solution to the solar neutrino problem is based upon an
incorrect assumption made in extrapolating nuclear cross sections
and the selective use of a small fraction of the nuclear physics
and of the neutrino data.  In addition, five different solar model
codes show that the rate obtained for the chlorine experiment using
the Dar-Shaviv stated parameters differs by at least $14 \sigma$
from the observed rate."}
\bigskip
\noindent
{\bf DS Response}:
\smallskip
\noindent
Dar and Shaviv did not claim to have solved the Cl solar
neutrino problem. Rather, they claimed agreement, {\bf within
theoretical  and experimental uncertainties}, between the
predictions of an improved standard solar
model and all solar neutrino observations {\bf after 1986}
(i.e., with all the published results of Kamiokande, SAGE and
GALLEX and with the Homestake observations {\bf after 1986)}.
This fact strongly suggests a standard physics solution to the solar
neutrino problem.
\smallskip
\noindent
Dar and Shaviv did consider  all relevant
nuclear data and did not introduce incorrect assumption in
extrapolating nuclear cross sections to low energies (see below).
\smallskip
\noindent
The five different SSM codes quoted by Bahcall et al.
do not and should not reproduce the results of Dar and Shaviv
when using their derived reaction rates for only two nuclear
reactions
and their value for the solar luminosity (Particle Data Group 1992):
Although Dar and Shaviv have stated explicitly
in their paper that for all the reactions, except for
$^7$Be(p,$\gamma)^8$B and $^3$He($\alpha,\gamma)^7$Be, they have
used the reactions rates compiled by Caughlan and Fowler (1988),
Bahcall et al. chose to "reproduce" the results of Dar and Shaviv
with different reaction rates. Moreover, the Bahcall et al.
codes include incorrect physical effects
and unjustified physical approximations which were not used
in the Dar-Shaviv calculations. In particular, these codes include
incorrect screening enhancement of nuclear reaction rates
in the Sun (omitted in the Dar-Shaviv code), assume nuclear
equilibrium for the CNO bi cycle (not assumed in the Dar-Shaviv
code), impose complete ionization of all elements
everywhere in the Sun in calculating diffusion and the equation of
state (not imposed in the Dar-Shaviv SSM code),
replace a complex variety of heavy
elements by a single effective element in calculating diffusion,
opacities and the equation of state (80 different isotopes are
included explicitly in the Dar-Shaviv SSM code. In particular, the
Dar-Shaviv code follows the destruction of the light elements from
Pre Main Sequence composition till the present day)
and do not calculate the evolution
of the Sun  during the Pre Main Sequence stage (the Dar-Shaviv SSM
calculations  include  the  evolution of the Sun during the Hayashi
phase). {\bf Most surprisingly,
Bahcall et al. have ignored the fact that the Bahcall-Pinsonneault
results were reproduced quite accurately by the code used by Dar
and Shaviv when similar physical assumptions and
approximations were used, i.e. before introducing the above
improvements (see Kovetz and Shaviv 1994).}

\smallskip
\noindent
In their paper Dar and Shaviv have stated explicitly that
the results of the Homestake experiment prior to 1986 are
inconsistent with their improved SSM predictions but they
also pointed out that these results look statistically incompatible
with the Homestake results after 1986, assuming a constant Sun.
The $14\sigma$ discrepancy, between the Bahcall et al. "reproduction"
of the Dar-Shaviv prediction and the observed rate in the Cl
experiment averaged over 23 years, is partly due to improper
reproduction of the Dar-Shaviv results by Bahcall et al. and
partly due to not including in $\sigma$ the large theoretical
uncertainty and the most probably large unknown systematic
errors in the Cl results (as implied  by the strong variation
of the results as function of time).

\smallskip
\noindent
The Homestake results after 1986 are compatable with
those obtained by Kamiokande during the same time period.
Furthermore,
the results of GALLEX and SAGE can be predicted correctly
from these observations and the solar luminosity using essentially
only standard physics conservation laws (shown in the Dar-Shaviv
paper). Therefore, Dar and Shaviv concluded that the solar neutrino
experiments do not provide evidence for physics beyond the standard
electroweak model and that only future
experiments like SNO and Superkamiokande will be able to
rule out a standard physics solution to the solar neutrino problem.

\bigskip
\noindent
$ \underline{Bahcall \ et \ al. \ claim:}$
\noindent
\smallskip
\noindent
{\it "2. Choosing Part of the Chlorine Solar-Neutrino Experimental
Data }

\bigskip
\noindent
{\it Dar and Shaviv chose to consider (see their Figure~1) only
four years of data from the chlorine solar neutrino experiment
beginning in 1987, although 23 years of data have been
reported. For the period beginning in 1987, the experimental
measurement of the chlorine rate is $2.8 \pm 0.3$ SNU. The
measured rate for the entire period for which data has been
reported is $2.28 \pm 0.23$ SNU. The measured rate in the
chlorine experiment, during the short period considered by
Dar and Shaviv, is still more than $4 \sigma$ less than their
calculated result of $4.2$ SNU and is more than $14 \sigma$
from the value we calculate for their stated parameters."}

\bigskip
\noindent
{\bf DS Response}:
\smallskip
\noindent
Dar and Shaviv stated clearly in their paper (in the abstract,
text and conclusions) that they compared their predictions with
{\bf all} the reported solar neutrino observations after 1986, i.e.,
with {\bf all }the reported results from Kamiokande, SAGE and
GALLEX and from runs  90-120 of the Homestake Cl experiment
during the time period 1986.8-1992.4, and not with
 ``only four years of data''. In Fig. 1 they showed
the data published by Homestake and Kamiokande for
only the period 1987-1990 in order to avoid showing
experimental data that was privately communicated to them
by the authors prior to their publication.

\smallskip
\noindent
The $^{37}$Ar average production rate observed in the Cl experiment
in runs  90-120 (during the time period 1986.8-1992.4), which
has been announced in recent international conferences and
has been communicated to us also by K. Lande
(who is one of the co-authors of the paper by Bahcall et al.),
is $0.61\pm 0.5$ atom/day. Since  1 atom/day = 5.35 SNU,
this rate corresponds to $3.26\pm 0.28$ SNU
before background subtraction and to $3.01\pm 0.28$ SNU
after background subtraction of $0.047\pm 0.016$ atom/day.
The last number is the present best estimate
of the background in the Cl experiment,
reported by Ray Davis at two recent international meetings:
"Neutrino Telescopes 1994" in Venice  and "The Solar Neutrino
Problem - Astrophysical  or Particle Physics Solution" in Grand Sasso.
\smallskip
\noindent
The average production rate of $^{37}$Ar in Cl during the time
period 1970.8 - 1985.3 observed by the Homestake experiment, as
was communicated to us by K. Lande, is $0.47\pm 0.04 \ $atoms/day
which corresponds to $2.09\pm 0.27 $SNU after $0.08\pm0.03 \
$atoms/day background subtraction, or to $2.26\pm 0.023 \ $SNU
after $0.047\pm 0.016 \ $atoms/day  background subtraction.

\smallskip
\noindent
As explained in our paper, the Homestake results prior and
after 1986 look statistically incompatible with a steady state Sun.
The annual $^{37}$Ar production rate has increased by more than
50\% since the beginning of the Homestake experiment. One can
ignore that and compare theory with the rate averaged over the
whole live time of the Homestake experiment, assuming
that the difference between the results before and after 1986
is due to a statistical fluctuation (with a rather small
probability) as advocated by Bahcall et al. and by
some other authors. Our approach is different.
Kamiokande began its Solar Neutrino observations
in 1987. It is only natural to compare the results of the
Homestake and Kamiokande experiments
during the time period when both were running, in particular,
in view of the fact that Ray Davis and other authors proposed
that solar neutrino flux is time dependent and anticorrelates with
the solar activity. It is worth to note, and perhaps even significant,
that after the installment of the new pumps in the Cl experiment
in 1986 the results from the new Cl experiment
(Homestake II, runs 90-120 during 1986.8-1992.4)
are not in conflict with the results from Kamiokande. Moreover,
during this time period neither the Homestake results
nor the Kamiokande results showed any statistically significant
anticorrelation with solar activity, nor any significant
time variation, which have been claimed with high
significance for the Cl data prior to 1986
by a few authors of the Bahcall et al. paper and by other authors.

\smallskip
\noindent
In view of all these facts and in view of the agreement between our
SSM predictions and the Kamiokande results for the $^8$B solar
neutrinos, it is scientifically unjustified to overlook the possibility
that the results after 1986 are closer to the truth than both (a) the
results before 1986 and (b) the weighted average of the results
before and after 1986.
\smallskip
\noindent
The Dar-Shaviv prediction for the Cl experiment is
$4.2\pm 0.6 \ $SNU. The results from the uncalibrated
Cl experiment after 1986 correspond to $3.01\pm 0.28$
SNU, where the error is statistical only, not including possible
systematic errors. Bahcall et al. concluded that the experimental
result is $4\sigma$ below the theoretical prediction ignoring
the theoretical uncertainty and possible systematic errors.
We, however, consider the above results to be consistent
within the theoretical and experimental uncertainties.
\smallskip
\noindent
The impressive $14\sigma$ discrepancy between the expected
rate in the Cl experiment, as calculated by Bahcall et al.
with the Dar-Shaviv nuclear parameters, and the average rate
at Homestake obtained for the whole 23 years of observations
is partly due to the incorrect physical approximations used
in the Bahcall et al. calculations (see also below),
partly due to ignoring the large uncertainty in the theoretical
prediction, and perhaps due to possible unknown large
systematic errors in the uncalibrated Cl experiment
(as implied by the significant increase of the rate as a
function of time since the starting of the Cl experiment).
\bigskip
\noindent
 {\bf $\underline{Bahcall \ et \ al. \ claim.}$
\smallskip
\noindent
{\it "3. Extrapolating Nuclear Cross Sections with an ad hoc
Assumption and Using Only Part of the Experimental Data }}

\bigskip
\noindent
{\it The theoretical models that have been used previously
{Johnson et al. 1992, Kajino et al. 1984, Parker and Rolfs 1991}
to extrapolate the cross section data to solar energies
properly take account of the finite  nuclear size effects
along with the other effects .....
These effects include nuclear structure, the
strong interaction, energy dependent operators in the transition
matrix elements, antisymmetrization between the colliding nucleons,
finite nuclear size, the final-state phase space, and the
contributions from other partial waves....
\smallskip
\noindent
... Dar and Shaviv assumed incorrectly that the only energy-
dependent
effect besides point-nuclei barrier penetration is nuclear size...
Dar and Shaviv did not take account of all the energy
dependencies nor of all the available nuclear physics data...
Their selective use of data and this incorrect assumption explain
why the Dar and Shaviv answers for $S(0)$ differs from the
standard values obtained by nuclear physicists ..."}

\bigskip
\noindent
{\bf DS Response:}
\smallskip
\noindent
 Bahcall et al. describe a wishful idealistic situation, where
 nuclear reaction theory is an exact and tested theory which
correctly predicts all cross sections (or at least
their energy dependence over a wide range).
This situation is far from reality. Actually,
nuclear reaction theory is an approximate theory which
generally provides only a reasonable parametrization
of measured nuclear cross sections over a limited range
of energy using many free parameters that are directly
adjusted to fit the data. In fact, for most reactions, including
the major reactions in the Sun,
it cannot even distinguish which of the different measurements
of a cross section is the correct one (normalization as well
as energy dependence). To make our point more concrete, let
us consider the model of Johnson et al. (1992)) used
"by the nuclear physicists" (together with an ad hoc procedure)
to extrapolate the various measurements of the cross section for
$^7$Be(p,$\gamma)^8$B to zero energy to obtain
S17(0)=22.4  eV-b, the value advocated by Bahcall et al.
In spite of all the good features of that model, which were
listed by Bahcall et al., Johnson et al. (1992) have not
demonstrated that their model correctly predicts:

\smallskip
(i) the p-wave resonance (i.e. its position, width and  magnitude),

\smallskip
(ii) the magnitude of the cross section,

\smallskip
(iii) the energy dependence of the cross section over the entire
energy range where it has been measured.
\bigskip
\smallskip
\noindent
In fact Johnson et al. did not trust their model energy
dependence (which they refer to as "suspect theoretical
calculations above 430 KeV" cf.  the last paragraph in Ap.J.
392,320,(1992) page 325) to extrapolate the
cross sections measured above the resonance energy by
Kavanagh 1960, Parker 1966, 1968 and Vaughn et al. 1970, and
choose instead to extrapolate these data to low energies below the
resonance according to the energy dependence of the averaged
Kavanagh et al. 1969 and Filiponne et al. 1983  experiments (an ad
hoc prescription).

\smallskip
\noindent
Moreover,

\smallskip
(iv) for the $^7$Li(n,$\gamma)^8$Li
capture cross section Johnson et al. predicted 30.6 mb while the
experimental values are $40.2\pm 2~ mb$ (Imhof et al. 1959) and
$45\pm 3.0~mb$ (Lynn et al. 1991),

\smallskip
(v) for the I=2 $^7$Li+n isotopic analog channel
Johnson et al. predicted a scattering length $a_2=0.26~fm$, while
the experimental  value is $a_2=-3.59\pm 0.06~fm$ .
\smallskip
This situation is rather typical (see for instance Descouvemont
and Baye 1994, in particular Figs. 5.6. therein).
\bigskip
\noindent
The above points are not intended to criticize the
otherwise good  works of Johnson et al., and
Descouvemont and Bay, but rather to
emphasize our contention that since no exact theory exists
for either the complicated quantum mechanical nuclear many
body problem or the direct nuclear
reactions,  one should try to rely as much as possible on "model
independent" features rather than on the choice of a specific
model for extrapolating cross sections to very low energies.
\bigskip
\noindent
Three such features were used in our paper:

\smallskip
(1) The measured position, width and height of a low energy
resonance in a given partial wave together with
the effective range approximation determine uniquely
the contribution of that partial wave to the total cross section all
the way down to zero energy (Kim et al., 1994). It was used by us to
subtract the p-wave contribution from the cross section for
$^7$Be(p,$\gamma)^8$B.

\smallskip
(2) Because of the Coulomb and centrifugal barriers, the
relative contribution of different non resonating partial waves
to the total cross section, at energies well below the Coulomb
barrier, are almost model independent, while the magnitude
and energy dependence of the individual partial waves are model
dependent. Thus, we used only the ratios calculated by various
groups (but neither the magnitude nor the energy dependence) to
extract from the measured cross section the s-wave contribution.

\smallskip
(3) Bound state wave functions decrease exponentially
outside the nucleus while the incident Coulomb wave functions for
energies well below the Coulomb barrier decrease
exponentially from the classical turning point
towards the nuclear surface. Consequently, most of the
contribution to the overlap integrals comes from the vicinity
of the nuclear surface. The s-wave cross section therefore,
is proportional to the absolute magnitude squared of the
incident Coulomb wave function near the nuclear surface,
rather than at the origin, i.e. to the Gamow (WKB) barrier
penetration factor for finite nuclear radii rather than
the Gamow barrier penetration factor for point nuclei.

\smallskip
\noindent
When extracting this energy dependence from the s-wave
contributions to the measured cross sections for
the three  reactions, $^7$Be(p,$\gamma)^8$B, $^3$He($\alpha,
\gamma)^7$Be and $^3$He($^3$He,2p)$^4$He, we found no evidence
for any significant energy dependence of our defined $\bar S(E)$ for
E well below the Coulomb barrier.
The absence of any noticeable energy dependence in $\bar S(E)$
extracted  from the measured cross sections for the above three
reactions, while S(E) shows strong
energy dependence, which is not completely accounted for by the
complicated theoretical models, is in our opinion,
a further support for our simple extrapolation procedure.
\bigskip
\noindent
   $\underline {Bahcall \ et \ al. \ claim:}$
\smallskip
\noindent
{\it "For the determination of the low-energy cross section
factor, S$_{34}$, for the $^3He(\alpha,\gamma)^7Be$ reaction,
Dar and Shaviv apparently adjusted the radius
parameter $R$ so
that the energy dependence of ${\bar S}(E)$ is mostly removed for
two of the nine (Parker and Rolfs 1991) existing experiments.
(They seem not to have noticed that the value of $R$ that they obtain
is very different from the measured radius of $2.8$ fm determined
by electron scattering.) They did not allow for the
other energy dependencies discussed above
and they only took account of two of the experiments."}
\bigskip
\smallskip
\noindent
{\bf {DS Response} }:
\smallskip
\noindent
Dar and Shaviv did not adjust the radius parameter $R=
R(^3$He)+$R(^4$He) so "that the energy dependence of ${\bar S}(E)$
is mostly removed for two of the nine existing experiments".
Dar and Shaviv used $R=R(^3$He)+$R(^4$He)$\approx 3.6 \ fm$
where
$R(^3$He)$\approx 2.0 \ fm$ and $R(^4$He)$\approx 1.6 \ fm$  were
determined from electron scattering measurements (e.g., Strueve et
al. 1992; Wu et al.1994)
and from scattering of high energy strongly interacting particles
(e.g. Tanihata et al. 1988), respectively, to calculate
$\bar S_{37}(E)$ for all nine experiments (the world data). The
straight line in Fig. 2 represents  the weighted average  of
$\bar S_{37}(E)$ for these world data. (Dar and Shaviv
were aware that $R(^7$Be$) \ \approx  2.8 \  fm$, but
that is not the appropriate radius to be used).
\bigskip
\noindent
{ $\underline{Bahcall \ et \ al.\ claim:}$}
\smallskip
\noindent
{\it "Dar and Shaviv cited the preliminary Coulomb
dissociation work described in preprint form (Motobayashi et al.
1994) as evidence for a lower-than-standard value for the crucial
cross section factor for the $^7Be(p,\gamma)^8B$ reaction. When
the $E2$ contribution to this reaction is taken into account
(Langanke and Shoppa 94), the preliminary Coulomb-dissociation
value differs from the six direct measurements (Parker and
Rolfs 1991) of the $^7Be(p,\gamma)^8B$ cross section
by a factor of two while the estimated uncertainty
(Johnson et al. 1992) in direct measurements is only $11\%$.
Moreover, there are still some unanswered questions about
the application of the Coulomb-dissociation method for determining
radiative capture cross sections, aside from the experimental
difficulties inherent in covering a sufficient range in energy
and angle to validate the reliability of any inferences."}

\bigskip
\noindent
{\bf DS Response:}
\smallskip
\noindent
The results described in the paper by Motobayashi et al. (1994)
that was submitted for publication in Physical Review
Letters and was cited by Dar and Shaviv were shown (Fig. 3 of
Dar and Shaviv ) to be in good agreement with the "world average"
s-wave cross section that was extracted from the four measurements
of the $^7Be(p,\gamma)^8B$ reaction in the same energy range,
(which is smaller there by approximately a factor two than the
total cross
section)
and in particular with that extracted from Vaughn (1970) and
Filippone (1983) if one allows a small $(<10\%)$ M1
contribution. The world data that includes that of
Motobayashi et al. ( 1994) yield $S17(0)=17 \pm 2~eV~b $.
The data of Motobayashi et al. yields approximately the same value.
\smallskip
\noindent
Neither the suggestion that the results of Motobayashi et al.
(1994) are preliminary nor their  analysis by Langanke and
Shoppa (1994) justify their omission
from the "world data" that is used to determine S17(E). In fact,
the analysis of Langanke and Shoppa of the Motobayashi et al.
data is highly questionable. They used as input E2 nuclear  matrix
elements calculated by others (Kim, Park and Kim, 1987). But if
instead one
uses the E2 nuclear matrix element calculated by Descouvemenot
and Baye (1994) one gets an effect that is approximately 2.4
times smaller at 0.6 MeV.  In addition, Langanke and Shoppa
find the largest E2 contribution at 0.6 MeV, which is on the
M1 resonance. The E2 contribution of the resonance is of no direct
relevance to the low energy (s-wave) cross section.

In the data points of Motobayashi et al. (1994),
measured at 0.8 and 1.0 MeV, that Langanke and Shoppa analyzed,
there is essentially no evidence for an E2 component, as they
state themselves. In fact, in all 15 or so
data points shown by Motobayashi et al. (1994)
there is only one point at approximately ${4}^{0}$,
0.6 MeV bin (on resonance) where the data deviates by
approximately 2 $\sigma$ from a predicted pure E1 behavior
(the E2 angular dependence is
sufficiently different from the E1's). A claim based on that one
point (out of 15 shown and some 30 measured in total [Motobayashi
et at. private communication] that there is evidence for E2
contribution in the Motobayashi et al. data is a bit too much.
\bigskip
\noindent
 ${\underline{Bahcall \ et \ al. \ claim:}}$
\smallskip
\noindent
{\it "For other nuclear reactions, Dar and Shaviv have used
the low-energy cross section factors from an
earlier review (Caughlan  and Fowler 1988) which provided fitting
formulae suitable for use at temperatures ($\sim 10^9$ K)
much higher than are reached in the Sun ($\sim 10^7$ K).
The quantitative effect of these approximations is
difficult to estimate, especially since other authors
(see section 4) use--for solar
calculations--explicit formulae that are suitable for the lower solar
temperatures.  However, an approximate discussion of using the
fitting formulae for higher temperatures
at solar temperatures has been given (Bahcall 1992);
the principal effects of the high-temperature formulae are in
the direction to decrease the  predicted $^7$Be and $^8$B
neutrino fluxes."}

\bigskip
\noindent
{\bf DS Response:}
\smallskip
\noindent
 We have verified that for solar temperatures the reaction
rates for all the reactions, except
$^7$Be(p,$\gamma)^8$B and $^3$He($\alpha,\gamma)^7$Be,
which were
compiled by Caughlan and Fowler (1988) are not significantly
different from those obtained by Bahcall and Ulrich (1989) and
by us. In fact the Kovetz-Shaviv
SSM code with the Caughlan - Fowler (1988) nuclear reaction rates
(for all reactions)
and all other assumptions and input as used by Bahcall and
Pinsonneault 1992 reproduces their predicted pp, pep, hep, $^7$Be
and $^8$B solar neutrino fluxes within $2\%$ accuracy
(see column KS of Table I of the Dar-Shaviv paper or
Kovetz and Shaviv 1994).
\bigskip
\noindent
 ${  \underline{Bahcall \ et \ al. \ comment:}}$
\smallskip
\noindent
{\it "Dar and Shaviv discuss at some length the fact
that the Debye-H\"uckel approximation to the screened nuclear
potential is not correct everywhere in the Sun.  It is not
clear what they recommend (although they say the effect is small),
nor if they are aware that modern screening calculations go
well beyond what they discuss..."}
\bigskip
\noindent
{\bf DS Response:}
\smallskip
\noindent
Dar and Shaviv discuss at some length the reasons why the
screening enhancement of the nuclear reaction rates near the center
of the Sun are negligible. The conventional screening enhancement
factors used by Bahcall et al. in their  SSM codes should be
taken out. This modification in the Bahcall-Pinsonneault
SSM code will reduce significantly their predicted $^8$B
solar neutrino flux (about $15 \%$). It does not change
significantly the predictions if this modification is introduced
after implementing all our other suggested improvements.
\bigskip
\noindent
 $ {\underline{Bahcall \ et \ al.}}$
\smallskip
\noindent
{\it 4. Solar \ Model \ Calculations}

\smallskip
\noindent
{\it Six authors of this paper (Bahcall, Christensen-Dalsgaard,
Degl'Innocenti, Glasner, Pinsonneault, and Proffitt)
have repeated the solar model calculations of Dar and Shaviv using
the non-standard parameters that Dar and Shaviv chose, namely, a
solar luminosity of $3.826 \times 10^{33}~{\rm erg~s^{-1} }$ and low
energy cross-section factors of $S_{34}(0)~=~0.45$ KeV-b and
$S_{17}(0)~=~17$ eV-b. Dar and Shaviv did not specify in their
preprint many of the important input quantities in their model;
they did not state what they used for the element abundances,
the radiative opacities, the equation of state,
and the neutrino cross sections.  They did not say which of the
several available prescriptions for diffusion they used.
We have therefore carried out calculations using a variety of
different choices for these quantities, namely, the choices
made previously as their best estimates by the six different
authors who used five independent stellar evolution codes} ...
{\it None of the well-tested solar codes that we have used
are able to reproduce the Dar and Shaviv results for neutrino
fluxes.}

\bigskip
\noindent
{\bf DS Response}:
\smallskip
\noindent
The input physics used by Dar and Shaviv is the most updated one.
The detailed SSM code with all input physics which was used by
Dar and Shaviv is documented in detail in the paper of Kovetz
and Shaviv (Ap. J, May 1, 1994) which is cited  by Dar and
Shaviv. The Kovetz-Shaviv paper is and was available in a preprint
form upon request. Bahcall et al. apparently overlooked this paper
and many physical improvements introduced by Dar and Shaviv
in the SSM calculations. All the five different SSM
codes of Bahcall et al. do not and should not reproduce the
results of Dar and Shaviv when using the same measured
solar luminosity that was used by Dar and Shaviv and the same
nuclear reaction rates for {\bf two reactions only},
$^7$Be(p,$\gamma)^8$B and $^3$He($\alpha,\gamma)^7$Be,
because these SSM codes use reaction rates different
from the Caughlan-Fowler (1988) reaction rates used by Dar
and Shaviv for all other reactions,
and because they include incorrect physical effects and
unjustified approximations which were avoided in the Dar-Shaviv
calculations: They include incorrect screening enhancements of
nuclear reaction rates in the Sun (which were not included in
the Dar-Shaviv calculations), assume (some of them) nuclear
equilibrium for the CNO bi cycle (which was not assumed by Dar
and Shaviv), impose complete ionization of all elements everywhere
in the Sun in calculating diffusion and the equation of state
(which was not imposed by Dar-Shaviv),
replace a complex variety of heavy elements by a single
effective element in calculating diffusion, opacities and
equations of state (while Dar-Shaviv considered separately
each of the 80 isotopes) and do not calculate the evolution of
the Sun  during the pre main sequence stage (while Dar and Shaviv
did). Most surprisingly, Bahcall et al. have ignored the fact
that the Bahcall-Pinsonneault results are reproduced quite
accurately by the code used by Dar and Shaviv when they imposed
the same physical approximations that were used by Bahcall and
Pinsonneault. (The Dar-Shaviv code also includes
several improvements and enhancements in the accuracy of
evolutionary models calculations never applied before).
\smallskip
In particular, Bahcall (1992) found the following empirical
dependence
of the $^8$B solar neutrino flux on the cross sections for the
major nuclear reactions in the Sun:
$$ \phi_{\nu_\odot}(^8{\rm B})\sim \sigma_{11}^{-2.6}
\sigma_{33}^{-0.4}\sigma_{34}^{0.8}\sigma_{17}^{1.0}~. $$
Thus, the use of our values (MeV-b units), $S_{11}=4.07\times
10^{-22}$, $S_{33}=5.60\times 10^3$ $S_{34}=0.45$ and $S_{17}=0.017$
instead of the values $S_{11}=4.00\times 10^{-22}$,
$S_{33}=5.0\times 10^3$, $S_{34}=0.533$ and $S_{17}=0.0224$
which were used by Bahcall et al. should decrease the Bahcall -
Pinsonneault (1992) prediction for $ \phi_{\nu_\odot}(^8{\rm B})$
by a factor $\approx 0.61~.$ Similarly, the omission of the weak
screening enhancement factors, $exp({U(0)/kT})$ where
$U(0)\approx Z_1Z_2e^2/R_D$, from the nuclear reaction rates
should decrease the $^8$B neutrino production rate
near the center of the Sun $(T_c\approx 1.571\times 10^7
{}~K$ and $R_D\approx
2.8\times10^{-9}cm)$ by a factor $\approx 0.89~.$ Thus, the combined
effect of the use of the Dar-Shaviv reaction rates and the omission of
the weak screening enhancement, according to Bahcall 1992, is
to reduce the Bahcall - Pinsonneault (1992) prediction,
$ \phi_{\nu_\odot}(^8{\rm B})\approx 5.69\times 10^6~cm^{-2}s^{-1}~,$
by a factor $\approx 0.54$, i.e.,
to $\phi_{\nu_\odot}(^8{\rm B})\approx 3.09\times 10^6~cm^{-2}s^{-1}$.
The reduced solar luminosity brings it down by additional $5\%$ and
all other improvements/changes introduced by Dar and Shaviv should
bring it further down (to their calculated value,
$ \phi_{\nu_\odot}(^8{\rm B})\approx 2.77\times 10^6~cm^{-2}s^{- 1}$ ?).

\smallskip
\noindent
In view of all these, we find no point in discussing the conclusions of
Bahcall et al. which are based on misleading comparisons.
Bahcall et al. should have introduced all the suggested
improvements/changes in their independent codes
and then try to reproduce the Dar-Shaviv results
before drawing their conclusions.
\smallskip
\noindent
\noindent
 ${  \underline{Bahcall \ et \ al.}} $
\smallskip
\noindent
{\bf {5. Conclusions:}}
\smallskip
\noindent
{\it `` Dar and Shaviv did not succeed in solving the solar neutrino
problem... Their failure to solve the problem is not surprising
since it has been demonstrated elsewhere (Bahcall and Bethe 1993)
that the chlorine and the Kamiokande experiments are inconsistent
another (if one assumes standard electroweak theory)...''}

\bigskip
\noindent
{\bf DS Response:}
\smallskip
\noindent
Dar and Shaviv did not claim to have solved the Cl solar
neutrino problem. They claimed agreement,
within theoretical and experimental uncertainties,
between the predictions of an improved standard solar
model and all solar neutrino observations  {\bf after 1986}
(i.e., with all the published results of Kamiokande,
SAGE and GALLEX and with the Homestake observations {\bf after
1986}). That suggest a standard physics solution to the
solar neutrino problem. Moreover,
the Homestake results after 1986 are consistent with
those obtained by Kamiokande during the same time period.
Their results together with the observed solar
luminosity predict correctly the observed production rate in
GALLEX and SAGE, using essentially only standard physics
conservation laws (see Dar and Shaviv 1994). Finally, the
spectrum of $^8$B neutrinos measured by Kamiokande is not
different from that predicted by the standard electroweak model.
In view of all these facts Dar
and Shaviv concluded that the solar neutrino experiments
do not provide solid evidence for physics beyond the standard
electroweak model. Only future experiments with large
statistics and precise measurements of the energy spectrum
and the lepton flavor content of solar neutrinos,
such as the Superkamiokande light water experiment
and the SNO heavy water experiment will be able to provide
reliable evidence for physics beyond the standard electroweak
model, which has not been found so far in
all precision tests of the standard electroweak model
at accelerators.

In fact, the standard solar model, which is only
an approximate description of the Sun, is surprisingly
successful in predicting the results of the pioneering solar
neutrino observations, in particular those obtained after 1986.
It is our considered opinion  that both the standard solar model
and the solar neutrino experimental results consist  a great
triumph for theoretical and experimental physics, in spite
of the so called "solar neutrino problem".

\vfill
\eject
\smallskip
\noindent
\centerline{{\bf Bibliography} }
\bigskip

\smallskip
\noindent
A.I. Abazov, Nucl. Phys. B (Proc. Supppl.) {\bf 19},
84 (1991).

\smallskip
\noindent
P. Anselmann, Phys. Lett. B {\bf 314}, 445 (1993).

\smallskip
\noindent
 P. Anselmann, et al., Phys. Lett., submitted, February (1994).

\smallskip
\noindent
J.N. Bahcall, and M. H. Pinsonneault, Rev.
Mod. Phys. {\bf 64}, 885 (1992).

\smallskip
\noindent
J.N. Bahcall, and H.A. Bethe, Phys. Rev. D
{\bf 47}, 1298 (1993).

\smallskip
\noindent
J.N. Bahcall, and M.H. Pinsonneault, in
preparation (1994).

\smallskip
\noindent
J.N. Bahcall, and A. Glasner, Ap. J., submitted
(1994).

\smallskip
\noindent
J.N. Bahcall, C.A. Barnes, J. Christensen-
Dalsgaard, B.T. Cleveland  S. Degl'innocenti,
B.W. Filippone, A. Glasner, R.W. Kavanagh, S.E. Koonin,
K. Lande, E.K. Langanke, P.D. Parker,
M.H. Pinsonneault,  C.R. Proffitt and T. Shoppa, Preprint 1994.

\smallskip
\noindent
C.A. Carraro, A. Schafer, and S.E. Koonin,
Ap. J. {\bf 331}, 565 (1988).

\smallskip
\noindent
V. Castellani, S. Degl'Innocenti, and G.
Fiorentini, A \& A. {\bf 271}, 601 (1993).

\smallskip
\noindent
G.R. Caughlan, and W.A. Fowler, At. Data Nucl. Data
Tables {\bf 40}, 283 (1988).

\smallskip
\noindent
J. Christensen-Dalsgaard, C.R. Proffitt, and
M.J. Thompson, Ap. J. {\bf 403}, L75 (1993).

\smallskip
\noindent
A. Dar, and G. Shaviv, Phys. Rev. Lett., preprint
submitted (1994).

\smallskip
\noindent
R. Davis Jr., in Frontiers of Neutrino
Astrophysics, ed. Y. Suzuki, and K. Nakamura (Tokyo: Universal
Academy Press, Inc., 1993), p. 47.

\smallskip
\noindent
P. Descouvemont and D. Baye, Nucl. Phys., {\bf A567}, 420 (1994).

\smallskip
\noindent
B.W. Filippone et al., Phys. Rev., {\bf C28}, 2222 (1993).

\smallskip
\noindent
W.A. Fowler, Rev. Mod. Phys. {\bf 56}, 149 (1984).

\smallskip
\noindent
G.M. Griffiths, M. Lal, and C.D. Scarfe,
Can. J. Phys. {\bf 41}, 724 (1963).

\smallskip
\noindent
W.L. Imhofet al., Phys. Rev. {\bf 114} 1037, 1959.

\smallskip
\noindent
C.W. Johnson, E. Kolbe, S.E. Koonin, and
K. Langanke, Ap. J. {\bf 392}, 320 (1992).

\smallskip
\noindent
T. Kajino and A. Arima, Phys. Rev. Lett. {\bf 52},
739 (1984).

\smallskip
\noindent
K. Langanke, and T. Shoppa (CalTech preprint), Phys. Rev.
C, in press (April, 1994).

\smallskip
\noindent
J.E. Lynn et al., Phys. Rev. {\bf C44}, 764 (1991).

\smallskip
\noindent
T. Motobayashi, et al., Phys. Rev. Lett.,
submitted (1994).

\smallskip
\noindent
R.W. Kavanagh, Nucl. Phys. {\bf 15}, 411 (1960).

\smallskip
\noindent
R.W. Kavanagh et al., Bull. Am. Phys. Soc. {\bf 14}, 1209 (1969).

\smallskip
\noindent
K.H. Kim et al., Phys. Rev. {\bf C35}, 363 (1987).

\smallskip
\noindent
Y.E. Kim et al., Purdue University preprint PNTG 94-2.

\smallskip
\noindent
A. Kovetz and G. Shaviv, Ap.J., May 1st 1994.

\smallskip
\noindent
P.D. Parker Phys. Rev. {\bf 150}, 851 (1966).

\smallskip
\noindent
P.D. Parker Ap. J. {\bf 153}, L85 (1968).

\smallskip
\noindent
P.D. Parker, and C. Rolfs, C. in The Solar
Interior and Atmosphere, ed. A. Cox, W. C. Livingston, and M. S.
Matthews
(Tucson: University of Arizona, 1991), p. 31

\smallskip
\noindent
Particle Data Group, Phys. Rev. {\bf D45}, III.2 (1992).

\smallskip
\noindent
C.R. Proffitt, Ap. J. {\bf 425}, 849 (1994).

\smallskip
\noindent
W. Strueve et al. Nucl. Phys., {\bf A537}, 367 (1992).

\smallskip
\noindent
Y. Suzuki, in Frontiers of Neutrino
Astrophysics, ed. Y. Suzuki, and K. Nakamura (Tokyo: Universal
Academy Press, Inc., 1993), p. 47.

\smallskip
\noindent
R. Tanihata et al. Phys. Lett., {\bf B206}, 592 (1998).

\smallskip
\noindent
F.J. Vaughn et al. Phys. Rev. {\bf C2}, 1657 (1970).

\smallskip
\noindent
R. D. Williams, and S. E. Koonin, Phys. Rev.
C {\bf 23}, 2773 (1981).

\smallskip
\noindent
Y. Wu et al. Few Body Systems, {\bf 15}, 145 (1993).

\end